\documentclass[twocolumn]{revtex4-1} 
\usepackage{graphicx,amsmath,amssymb}
\usepackage{mathrsfs,eufrak}
\usepackage{dcolumn}
\usepackage{bm}
\begin{document}
\title{Tilted Dirac Cones in Two-Dimensional Materials: Impact on Electron Transmission and Pseudospin Dynamics}
\author{Rasha Al-Marzoog$^1$}
\author{Ali Rezaei$^1$}
\author{Zahra Noorinejad$^2$}
\author{Mohsen Amini$^1$}
\author{Ebrahim Ghanbari-Adivi$^1$}\email{ghanbari@phys.ui.ac.ir}
\author{Seyed Akbar Jafari$^3$}
\affiliation{$1$ Faculty of Physics, University of Isfahan, Isfahan
81746-73441, Iran}

\affiliation{$2$ Department of Physics,  Islamic Azad University -
Shahreza Branch, Shahreza, Iran}

\affiliation{$3$ II. Physikalisches Institute, RWTH Aachen
University, 52074 Aachen, Germany}
\begin{abstract}
This study is devoted to the profound implications of tilted Dirac
cones on the quantum transport properties of two-dimensional~(2D)
Dirac materials. These materials, characterized by their linear
conic energy dispersions in the vicinity of Dirac points, exhibit
unique electronic behaviors, including the emulation of massless
Dirac fermions and the manifestation of relativistic phenomena such
as Klein tunneling. Expanding beyond the well-studied case of
graphene, the manuscript focuses on materials with tilted Dirac
cones, where the anisotropic and tilted nature of the cones
introduces additional complexity and richness to their electronic
properties. The investigation begins by considering a heterojunction
of 2D~Dirac materials, where electrons undergo quantum tunneling
between regions with upright and tilted Dirac cones.  The role of
tilt in characterizing the transmission of electrons across these
interfaces is thoroughly examined, shedding light on the influence
of the tilt parameter on the transmission probability and the fate
of the pseudospin of the Dirac electrons, particularly upon a sudden
change in the tilting. We also investigate the probability of
reflection and transmission from an intermediate slab with arbitrary
subcritical tilt, focusing on the behavior of electron transmission
across regions with varying Dirac cone tilts. The study demonstrates
that for certain thicknesses of the middle slab, the transmission
probability is equal to unity, and both reflection and transmission
exhibit periodic behavior with respect to the slab thickness.
\end{abstract}
\maketitle
\section{Introduction~\label{Sec.I}}
The discovery of Dirac-like fermions in graphene has significantly
expanded the horizons of condensed matter physics and led to the
development of Dirac-fermion systems~\cite{Novoselov01}. These
materials exhibit remarkable electronic properties, characterized by
the presence of linear conic energy dispersions near the Dirac
points within their momentum space~\cite{Neto01}. Due to the linear
nature of the energy dispersion, the low-energy charge carriers
within these materials behave like massless Dirac fermions,
resulting in the emergence of relativistic
behavior~\cite{Neto01}.\par
The electronic properties of graphene can be controlled by applying
external electric and magnetic fields or altering sample geometry
and/or topology~\cite{Neto01}. Edge (surface) states in graphene
depend on the edge termination (zigzag or armchair) and affect the
physical properties of nanoribbons. Different types of disorder
modify the Dirac equation, leading to unusual spectroscopic and
transport properties~\cite{Neto01}. Also, the electron-electron and
electron-phonon interactions in single- and multi-layer 2D~materials
affect the behavior of Dirac fermions in these
materials.~\cite{Neto01}.\par
One of the most notable phenomena associated with the relativistic
behavior of graphene is Klein tunneling, a fundamental quantum
mechanical phenomenon rooted in the principles of relativistic
physics~\cite{Dombey01,Katsnelson01,Ghanbari01}. In graphene, the
low-energy excitations are massless, chiral Dirac fermions, and the
chemical potential crosses exactly the Dirac
point~~\cite{Katsnelson01}. This unique dispersion, valid only at
low energies, mimics the physics of quantum electrodynamics~(QED)
for massless fermions, except for the fact that in graphene, the
Dirac points occur at the vertices of the standard hexagonal
Brillouin zone~\cite{Katsnelson01}.\par
Dirac cones, can exhibit tilted dispersions along a preferred
direction of wave vector, introducing additional complexity and
richness to their electronic
properties~\cite{Liu01,Qian01,Lu01,Li01}. In materials with tilted
Dirac cones, the linear energy-momentum relation remains intact, but
the tilt in the cone results in anisotropic behaviors and opens up
exciting possibilities for tailoring their specific properties. This
tilt also naturally leads to an emergent spacetime
metric~\cite{Volovik01}.\par
Based on a parameter known as the tilt parameter,~$\zeta$, the
tilted Dirac cones can be classified into four primary
types~\cite{Volovik02,Volovik03,Soluyanov01}: Untilted type~($\zeta=
0$) for which the Dirac cone is not tilted, and its dispersion is
isotropic. Type-I or subcritical type~($0 < \zeta < 1$) for which
the Dirac cone is tilted, and its dispersion is anisotropic. This
type exhibits a range of anisotropic behaviors and can be tailored
to specific properties. Type-II or overcritical type ($\zeta > 1$)
for which the Dirac cone is highly tilted, and its dispersion is
strongly anisotropic. This type can lead to exotic phenomena such as
anomalous Hall effect or unconventional Klein
tunneling~\cite{Dombey01}. Type-III or critical type~($\zeta = 1$)
for which the Dirac cone is critically tilted, and its dispersion is
highly anisotropic. This type can exhibit intricate transport
phenomena, such as modified Klein tunneling~\cite{Dombey01}. In
summary, tilted Dirac cones in 2D~materials introduce additional
complexity and richness to their electronic properties, leading to
anisotropic behaviors and emergent spacetime metrics. These tilted
cones can be classified into four primary types based on the tilt
parameter, each exhibiting unique properties and potential for
tailoring specific applications.\par
The tilt observed in the energy dispersion cones of fermions has a
significant impact on various properties of these
particles~\cite{Yekta01,Goerbig01,Trescher01,Steiner01,Zhang01}. For
example, in 8-Pmmn borophene, the impact of tilted velocity and
anisotropic Dirac cones on phenomena like asymmetric and oblique
Klein tunneling and valley-dependent electron retroreflection has
been examined~\cite{Dombey01,Zhang01,Zhou01}. Variations in the tilt
magnitude lead to non-universal behavior in the anomalous Hall
conductivity~\cite{Zyuzin01}, with small~(subcritical) tilts
inducing an asymmetric Pauli blockade, resulting in finite
free-electron Hall responses~\cite{Steiner01} and non-zero
photocurrents~\cite{Ma01}, while large (overcritical) tilts cause
the Fermi surface to deviate from a point-like structure, resulting
in the emergence of a gap in the Landau-level spectrum and the
absence of chiral zero modes when the magnetic field direction falls
outside the cone~\cite{Soluyanov01}. Recent discussions have also
highlighted the considerable variation in optical conductivity among
different tilt types in 2D~materials with tilted Dirac
cones~\cite{Wild01,Tan01}. Furthermore, due to the anisotropy
associated with the tilting of the bands, these materials also
provide a rich solid-state space-time platform, marked by the
corresponding non-Minkowski metric~\cite{Mola01,Mola02}. Therefore,
it is fascinating and advantageous to investigate how the
characteristic tilting of the Dirac cones affects the transmission
of Dirac fermions across interfaces that connect regions with
varying types of tilted Dirac cones. This question becomes
particularly pertinent as active endeavors are underway to
manipulate and precisely tune the tilt in Dirac
materials~\cite{Goerbig01,Yekta01,Lu02,Long01}.\par
To comprehend the impact of the tilt in the Dirac cone, we
investigate the quantum transport in a heterojunction of 2D~Dirac
materials, where electrons tunnel between an upright Dirac cone and
a tilted Dirac cone. Our study focuses on the role of the tilt in
characterizing the transmission and the fate of the pseudospin of
the Dirac electrons upon a sudden change in the tilting. This
research is crucial for understanding the behavior of electron
transmission across regions with varying Dirac cone tilts, providing
valuable insights into quantum transport phenomena in such systems.
The investigation of the effects of Dirac cone tilt not only helps
us understand electron transmission probabilities through barriers
but also has significant implications for optimizing
resonant-tunneling quantum devices based on heterostructures, paving
the way for novel quantum device development.
This research is of significant interest due to the ongoing efforts
to manipulate and precisely tune the tilt in Dirac materials. The
manuscript also sheds light on the broader implications of tilted
Dirac cones, particularly in less symmetric Dirac materials, where
the Coulomb interaction can give rise to even more exotic phenomena.
Furthermore, the study delves into the theoretical exploration of
the excitonic transition in 2D~tilted cones to understand the
electron-hole pairing instability as a function of tilt, providing
valuable insights into the chiral excitonic instability of such
systems. In summary, the manuscript offers a comprehensive
theoretical investigation into the impact of tilted Dirac cones on
electron transmission and pseudospin dynamics in 2D~materials,
contributing valuable insights into these unique electronic systems
and their potential applications in quantum transport and beyond.
The study also undertakes a theoretical investigation into the
transmission characteristics of Dirac fermions across interfaces
linking regions with differing types of energy dispersion tilt. The
rest of the paper is organized as follows.\par
Section~\ref{Sec02} explains the desired model and its results. In
subsection~\ref{Sec02SubA}, we first introduce the Hamiltonian and
the eigenstates of the tilted Dirac cone. Subsequently, in
subsection~\ref{Sec02SubB}, based on the continuity equation, we
derive the appropriate boundary conditions between two regions with
different tilts and calculate the spin rotation at the interface. In
this subsection, we also determine the probability of electron
reflection and transmission from a non-tilted region to a tilted
region for both normal and oblique incidence. In
subsection~\ref{Sec02SubC}, we calculate the probability of
reflection and transmission from an intermediate slab with arbitrary
tilt. In particular, we have investigated the effect of slab
thickness on these possibilities. It is shown that by changing the
thickness of the slab, these probabilities will also change
periodically, and for certain thicknesses, one of them is equal to
unit while the other is zero. These analysis provide valuable
insights into the behavior of electron transmission across regions
with varying Dirac cone tilts, contributing to a deeper
understanding of quantum transport phenomena in such systems. A
summary along with the conclusion remarks are presented in
section~\ref{Sec03}.
\section{Theoretical formalism~\label{Sec02}}
In this section, we present the theoretical framework for examining
the behavior of Dirac fermions in 2D~tilted materials. We begin by
introducing the desired model and the corresponding Hamiltonian that
captures the unique characteristics of these fermions. Subsequently,
we determine the eigenvalues and eigenstates of the Hamiltonian to
gain insights into the fundamental properties of the system.
Following this, we investigate the behavior of fermions at the
interface of two materials, shedding light on their transmission
properties in this context. Additionally, we analyze the behavior of
fermions when they collide with a buffer slab of limited thickness
between two mediums. Finally, we thoroughly present and discuss the
results obtained from each case, providing a comprehensive
understanding of the behavior of Dirac fermions in the specified
scenarios.
\subsection{Hamiltonian Model for Dirac Fermions in 2D Tilted Materials~\label{Sec02SubA}}
The aim of this section is to construct and explain the theoretical
framework utilized in this research for examining the low-energy
transmission characteristics of tilted Dirac fermions. To this end,
we can start by examining the Hamiltonian that describes such a
process. The proper Hamiltonian is assumed as
\begin{equation}\label{Eq01}
{\mathcal{H}} = v_x  k_x \sigma_x + v_y  k_y \sigma_y+ v_t k_x {\bf
1},
\end{equation}
in which $k_x$ and $k_y$ are the electron momentum or wavevector
components in the $x$ and $y$ directions, respectively, $\sigma_x$
and $\sigma_y$ are the well-known Pauli matrices, and ${\bf 1}$ is a
$2 \times 2$ identity matrix. The model parameters $v_x$ and $v_y$
indicate the anisotropic velocities, while $v_t$ represents the tilt
velocity that represents \emph{nematicity} of the Dirac electrons
mingling space and time.\par
Introducing the tilt parameter as $\zeta = v_t /v_x$ allows us to
consider the band tilting along the $x$-direction. The value of this
parameter classifies the Dirac materials into four distinct phases.
$\zeta = 0$ indicates the untilted Dirac materials while $\zeta=1$
corresponds the critical tilting. For values of $\zeta$ between
$0$~and~$1$, the 2D~materials are categorized in subcritical class
and for values greater than $1$, they belong to overcritical
class.\par
For the sake of simplicity in the computations, and without loss of
generality, in the following, we assume that $v_x=v_y=v$. With this
assumption, the eigenvalues and eigenvectors for both the conduction
and valence bands can be easily obtained by diagonalizing the
Hamiltonian given in Eq.~\eqref{Eq01}. The obtained results are
\begin{equation}\label{Eq02}
E_{\lambda , {\bf k}} = v_t k_x + \lambda v k,
\end{equation}
and
\begin{equation}\label{Eq03}
\psi_{\lambda,{\bf k}}({\bf r})= {1 \over 2} \Big(
\begin{array}{c}
1\\ \lambda e^{i\varphi}
\end{array} \Big) e^{i {\bf k} \cdot \bf{r}},
\end{equation}
where ${\bf r} = (x, y)$ represents the position vector, ${\bf k} =
(k_x, k_y)= (k \cos\varphi, k\sin\varphi)$ denotes the wavevector
measured from the Dirac point, $k$ is the norm of the wavevector,
$\varphi =\tan^{-1}(ky/k_x)$, and $\lambda = \pm 1$ indicates the
conduction $(+1)$ and valence $(-1)$ bands, respectively. As is seen
from Eq.~\eqref{Eq03}, the directions of the pseudospin, ${\bf S}$,
and the wave vector, ${\bf k}$, are the same.
\subsection{Single interface heterojunction\label{Sec02SubB}}
The investigation of the behavior of Dirac fermions at the interface
of two 2D~Dirac materials is a pivotal step in our theoretical
framework. This analysis involves the examination of the low-energy
transmission properties of tilted Dirac fermions, which can be
described by the Hamiltonian given by Eq.~\eqref{Eq01}.\par
\begin{figure}[t!]
\begin{center}
\includegraphics[scale=0.65]{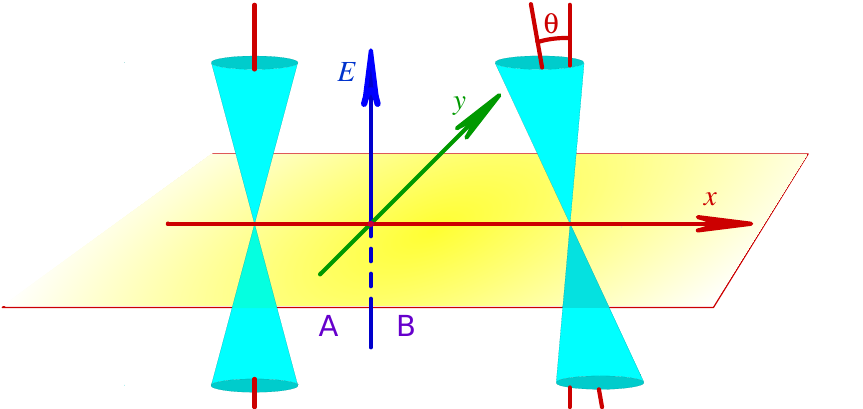}
\end{center}
\caption{A schematic representation of the behavior of electrons
when passing through a heterojunction of 2D~Dirac materials, where
electrons undergo quantum tunneling between regions with upright and
tilted Dirac cones.\label{Fig01}}
\end{figure}
In order to accomplish this task, we can begin our study by
considering a heterojunction formed by two different  Dirac
materials, each with distinct types of tilt in their energy bands,
as illustrated in Fig.~\ref{Fig01}. As depicted in the figure, the
mediums are labeled by $A$ and $B$, the Dirac cone on the left side
($x < 0$) remains untilted ($\zeta_A=0$), while it exhibits tilt
($\zeta_B\neq0$), on the right side ($x > 0$). Consequently, the
tilt velocity in Eq.~\eqref{Eq01} becomes a function of $x$ and can
be expressed as
\begin{equation}\label{Eq04}
v_t(x)=v_t\Theta(x)=v_t\Big\{\begin{array}{lr}0,\quad&\quad x<0,\\
1,\quad&\quad x>0,\end{array},
\end{equation}
where $\Theta(x)$ denotes the Heaviside unit step function.\par
It is important to highlight that in our investigation, we
specifically address the scenario with Dirac cones tilted in the
$k_x$-direction, which is perpendicular to the interface. This
configuration is distinct from cases where the tilt is parallel to
the interface, as discussed in the
literature~\cite{Pattrawutthiwong01}.\par
\begin{figure*}[t]
\begin{center}
\includegraphics[scale=0.5] {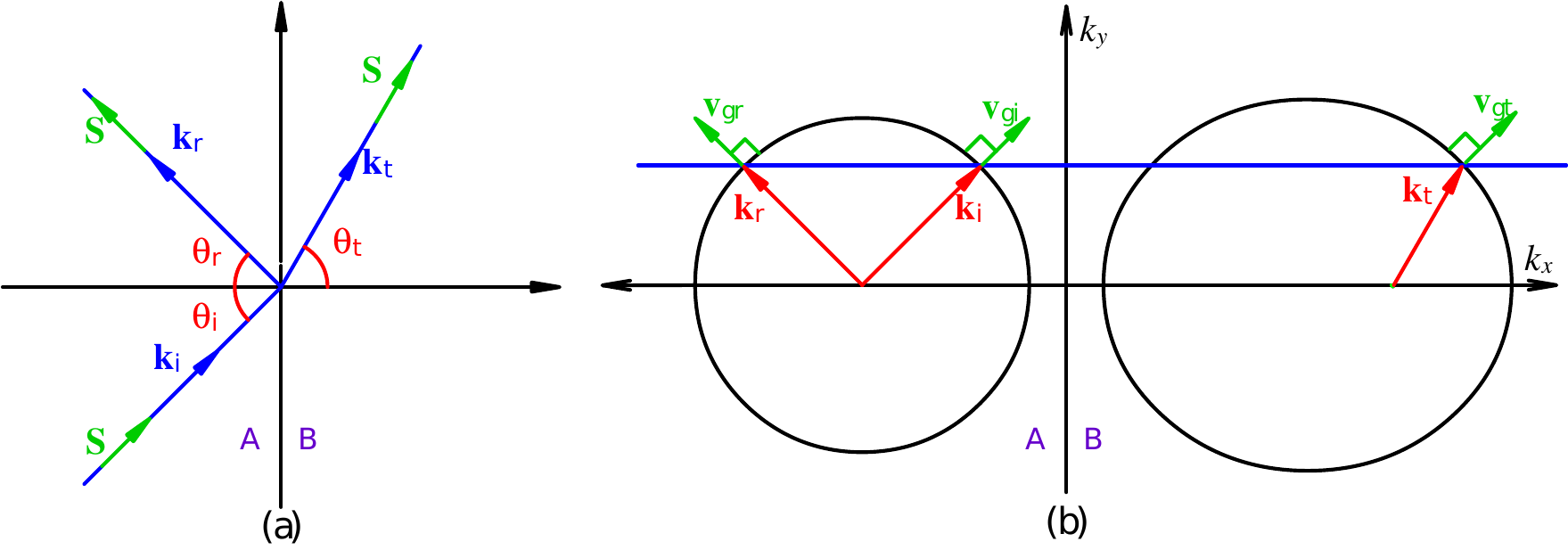}
\end{center}
\caption{Reflection and transmission of the electron waves at the
interface between two 2D~materials for an oblique incidence: (a)
Orientation of the wave vectors ${\bf k}_i$, ${\bf k}_r$ and ${\bf
k}_t$ the pseudo-spin~${\bf S}$ at the interface,~(b)~The
intersections of the Fermi surface and the Dirac cones in regions
$A$ and $B$. The Dirac cones for the second region are tilted. ${\bf
v}_{gi}$, ${\bf v}_{gr}$, and ${\bf v}_{gt}$ are the group
velocities of the incident, reflected and transmitted waves,
respectively. \label{Fig02}}
\end{figure*}
We are interested in understanding the transmission properties
across the interface between regions $A$ and $B$, as illustrated in
Fig.~\ref{Fig01}. Before delving into the calculations of reflection
and transmission coefficients, let us establish the boundary
conditions by utilizing the continuity equation. In elementary
quantum mechanics, it is well known that the probability density,
$\rho$, and the probability current density, $\bf{J}$, satisfy the
continuity equation:
\begin{equation} \label{Eq05}
{\partial \rho \over \partial t}+\bf{\nabla} \cdot \bf{J} = 0,
\end{equation}
representing the conservation of the probability in a given system
which is a fundamental concept in quantum mechanics. Now, assume
that the Hamitonian describing the transport of an untilted Dirac
fermion in a \emph{``hypothetical system''} is given by $H =
\sigma_z k_z = -i \sigma_z\partial/\partial z$. For this case, using
the time dependent Schr\"{o}dinger equation, it is easy to show that
the continuity equation is satisfied with $\rho$ and ${\bf J}$
defined in terms of wavefunction $\psi$ as $\rho = \psi^\dagger
\psi$ and ${\bf J} = (\psi^\dagger \sigma_z \psi){\bf e}_z$ in which
${\bf e}_z$ is the unit vector in $z$ direction. Also, in this case
it is an easy practice to show that the wavefunction $\psi$ is
continues everywhere even at the interface of two $2D$~Dirac mediums
with untilted Dirac cones. This fact implies that, at the interface
of these regions, we have $\psi_A=\psi_B$ where $\psi_A$ and
$\psi_B$ refer to the electron wavefunctions in regions $A$ and $B$
at the interface, respectively. It should be mentioned that $\psi$,
$\rho$ and $\bf{J}$ are all functions of the position $\bf{r}$ and
time $t$, but for the sake of brevity, explicit display of this
dependence has been avoided.\par
In similarity with what was stated above for the untilted Dirac
fermions, the continuity equation remains hold for the tilted Dirac
fermions obeying the Hamiltonian given in Eq.~\eqref{Eq01}, if ${\bf
J}$ is defined as ${\bf J}= J_x{\bf e}_x+J_y{\bf e}_y$ with $J_x=
\psi^\dagger (v_x\sigma_x + v_t{\bf 1}) \psi$ and $J_y=\psi^\dagger
v_y\sigma_y \psi$. Consequently, if $J_x$ can be written as
$J_x=\Psi^\dagger\sigma_z\Psi$, it can be concluded that $\Psi$ is
continues everywhere including the interface of the regions.
Specially, at the interface of the regions, we have:
\begin{equation}\label{Eq06}
\Psi_A = \Psi_B,
\end{equation}
which is a crucial point to continue the discussion.\par
With the above introduction we return to the main issue in question.
A schematic representation of the considered system is depicted in
Fig.~\ref{Fig02}. The figure illustrates two key aspects: (a) the
orientation of the wave vectors ${\bf k}_i$, ${\bf k}_r$, ${\bf
k}_t$, and the pseudo-spin ${\bf S}$ at the interface, and (b) the
intersections of the Fermi surface and the Dirac cones in regions
$A$ and $B$, with the cones in the second region being tilted. In
part (a), the orientation of the wave vectors and the pseudo-spin at
the interface is depicted, providing insight into the behavior of
the wave vectors and the associated pseudo-spin at the interface.
This orientation is crucial for understanding the dynamics of the
system at the interface. In part (b), the intersections of the Fermi
surface and the Dirac cones in regions $A$ and $B$ are shown, with
particular emphasis on the tilted nature of the cones in the second
region. This tilt has significant implications for the electronic
properties in this region, and the figure serves to visually convey
this important characteristic. The intersections of the Fermi
surface and the Dirac cones are fundamental to the electronic
structure and behavior of the material, and the tilt in the second
region introduces additional complexity to the electronic
properties, which is effectively captured in the figure.\par
If the probability current densities in regions $A$ and $B$ are
represented by $\bf{J}_A$ and $\bf{J}_B$, respectively, the
perpendicular components of these two vector quantities must be
equal at the junction  of the regions, that is: $J_{Ax} = J_{Bx}$.
This condition implies that
\begin{equation}\label{Eq07}
\psi_A^\dagger(v_A \sigma_x + v_{At} {\bf 1}) \psi_A =
\psi_B^\dagger(v_B \sigma_x+v_{Bt} {\bf 1}) \psi_B
\end{equation}
in which $v_A$ and $v_B$ are the $x$ components of the electron
velocities in regions $A$ and $B$, respectively.\par
A similar boundary condition has been derived for quantum transport
in the Weyl~semimetals of type~II in Refs.~\cite{Hashimoto01}
and~\cite{Witten01}. However, it is important to note two
differences. First, in the mentioned references, the right side of
the equation has been set equal to zero, which is due to the
assumption of a closed boundary condition in those studies, but in
the current research, the boundary condition is for the quantum
transport of the massless Dirac fermions from the interface of the
regions. The other is that in the mentioned studies, the similar
boundary condition has been obtained by providing some different and
a little complicated discussions, while in this research, the
appropriate boundary condition has been deduced simply by employing
the continuity condition.\par
Our attention is now directed towards expressing the continuity
condition given in Eq.~\eqref{Eq07} as
\begin{equation}\label{Eq08}
\Psi_A^\dagger\sigma_z \Psi_A = \Psi_B^\dagger\sigma_z \Psi_B,
\end{equation}
in order to establish the continuity of $\Psi$ at the boundary of
the regions as stated in Eq.~\eqref{Eq06}. For this purpose, using a
unitary transformation represented by the unitary operator of $U$
denoted as
\begin{equation}\label{Eq09}
U ={1\over \sqrt{2}}(\sigma_x+\sigma_z) = {1 \over \sqrt{2}} \Big(
\begin{array}{cc}
1 & 1 \\
1 & -1
\end{array}
\Big),
\end{equation}
it is possible to transform Eq.~\eqref{Eq07} to
\begin{equation}\label{Eq10}
\begin{split} v_A \psi_A^\dagger U^\dagger (\sigma_z & + \zeta_A {\bf 1}) U
\psi_A\\ & = v_B \psi_B^\dagger U^\dagger ( \sigma_z + \zeta_B {\bf
1}) U \psi_B,
\end{split}
\end{equation}
where $ \sigma_z = U^\dagger \sigma_x U$, and $\zeta_A$ and
$\zeta_B$ denote the tilt parameters for mediums $A$ and $B$,
correspondingly.
The boundary condition represented in Eq.~\eqref{Eq08} can be
expressed in its explicit matrix form as
\begin{equation} \label{Eq11}
v_A \psi_A^\dagger \Big (\begin{array}{cc}
\zeta_A & 1 \\
1 & \zeta_A
\end{array} \Big)
\psi_A = v_B \psi_B^\dagger \Big(
\begin{array}{cc}
\zeta_B & 1 \\
1 & \zeta_B
\end{array} \Big)
\psi_B.
\end{equation}
By converting this relation into the form given in Eq.~\eqref{Eq08}
and considering Eq.~\eqref{Eq06}, a continues quantity at the
interface is obtained such as:
\begin{equation} \label{Eq12}\begin{split}
\sqrt{v_A} \Big( \begin{array}{cc} \sqrt{1+\zeta_A} & 0 \\
0 & \sqrt{1-\zeta_A}\end{array} \Big) U \psi_A \qquad\qquad\qquad\qquad\\
=
\sqrt{v_B} \left( \begin{array}{cc} \sqrt{1+\zeta_B} & 0 \\
0 & \sqrt{1-\zeta_B} \end{array} \right) U\psi_B.\end{split}
\end{equation}
A similar expression has been also derived in
Ref.~\cite{Hashimoto01} for theoretical study of the escape from
black hole analogs in 2D~materials.\par
Ultimately, when both sides of Eq.~\eqref{Eq12} are multiplied from
the left by unitary matrix $U^\dagger$, the continuity condition of
that component of ${\bf J}$ which is perpendicular to the interface
at the junction of the regions is reduced accordingly as
\begin{equation} \label{Eq13}
M_A \psi_A = M_B \psi_B,
\end{equation}
where $M_A$ and $M_B$ are two $2\times 2$ matrices corresponding to
the mediums of $A$ and $B$, respectively. The explicit matrix form
of these matrices are given by
\begin{equation} \label{Eq14}
M_A  = \Big(
\begin{array}{cc}
a_A & b_A \\
b_A & a_A
\end{array}\Big),\quad {\rm and}\quad M_B  = \Big(
\begin{array}{cc}
a_B & b_B \\
b_B & a_B
\end{array}\Big)
\end{equation}
where
\begin{equation} \label{Eq15}
\begin{split}
a_A =
{1\over2}\sqrt{v_A}\Big(\sqrt{1+\zeta_A}+\sqrt{1-\zeta_A}\Big),\\
b_A = {1\over2}\sqrt{v_A}\Big(
\sqrt{1+\zeta_A}-\sqrt{1-\zeta_A}\Big).
\end{split}
\end{equation}
Similarly, $a_B$, and $b_B$ have the same forms as those in
Eq.~\eqref{Eq15}, except that index $A$ should be replaced
by~$B$.\par
It is evident that if the tilt velocities in both regions are equal,
i.e. $v_A = v_B$, it leads to $M_A = M_B$, resulting in the
simplification of the boundary condition of Eq.~\eqref{Eq13} to
$\psi_A = \psi_B$. This condition holds true even if $v_A = v_B =
0$, which indicates no tilt in the $k_x$-direction (though tilt may
still exist in the $k_y$-direction, as shown in
Ref.~\cite{Pattrawutthiwong01}). However, for our current research,
given the tilt in the direction perpendicular to the interface
between regions $A$ and $B$, we will utilize the boundary condition
provided in Eq.~\eqref{Eq13}, as we will discuss it further.\par
Prior to concluding this subsection, we address the rotation of the
``spin" or ``pseudospin" state of the Dirac fermions in confronting
with the interface of the regions at which a sudden change occurs in
tilt. For the sake of simplicity in the calculations, we revert to
the bases obtained after employing the unitary transformation
represented in Eq.~\eqref{Eq09}. In the specified bases, if the
``spin" or ``pseudospin"  states of the Dirac fermions, in regions
$A$ and $B$, are denoted by $\psi_{SA}$ and $\psi_{SB}$,
respectively, by referring to Eq.~\eqref{Eq12}, it is found that
\begin{equation} \label{Eq16}
\begin{split}
\psi_{SB} = \sqrt{v_A/v_B} & \Big (\begin{array}{cc}
\sqrt{1+\zeta_B} & 0 \\ 0 & \sqrt{1-\zeta_B} \end{array} \Big)^{-1}
\\ &\Big (\begin{array}{cc} \sqrt{1+\zeta_A} & 0 \\ 0 &
\sqrt{1-\zeta_A}
\end{array} \Big)\psi_{SA}.
\end{split}
\end{equation}
Using the basic calculus, the above relation can be easily reduced
to
\begin{equation}\label{Eq17}
\psi_{SB} = \Big (\begin{array}{cc} a_S & 0 \\ 0 & b_S
\end{array} \Big)\psi_{SA}.
\end{equation}
where $a_S$ and $b_S$ are given by the following expressions:
\[
a_S = \sqrt{v_A (1+\zeta_A)\over v_B(1+\zeta_B)},\quad {\rm
and}\quad b_S = \sqrt{v_A (1-\zeta_A)\over v_B(1-\zeta_B)}.
\]
We assume that the ``spin" or ``pseudospin" state $\psi_{SA}$  is
given in terms of the spherical coordinates by
\begin{equation} \label{Eq18}
\psi_{SA} = \begin{pmatrix}
\cos({\theta/2})\ e^{-i\phi/2}\\
\sin({\theta/2})\ e^{+i\phi/2}
\end{pmatrix}.
\end{equation}
This state is the eigenstate of ${\bf S}\cdot{\bf n}$ with
eigenvalue of $+1/2$. Here, ${\bf S}$ denotes the ``spin" or
``pseudospin" operator and ${\bf n}$ is an arbitrary unit vector
representing the direction associated with the polar and azimuthal
angles of $\theta$ and $\phi$, respectively. Consequently,
Eq.~\eqref{Eq17} leads to
\[ \psi_{SB}= \Big(\begin{array}{c}
a_S\cos(\theta/2)\ e^{-i\phi/2}\\
b_S\sin(\theta/2)\ e^{+i\phi/2}
\end{array}\Big).
\]
In order to examine the ``spin" or ``pseudospin" direction of
$\psi_{SB}$, the corresponding state can be rewritten as
\begin{equation}\label{Eq19}
\psi_{SB} = N \Big(\begin{array}{c}
\cos({\theta_S/2})\ e^{-i\phi/2}\\
\sin({\theta_S/2})\ e^{+i\phi/2}
\end{array}\Big),
\end{equation}
where $N$ is a normalization factor, and
\begin{equation}\label{Eq20}
\theta_S = 2 \tan^{-1}\big[b_S \tan(\theta/2)/ a_S\big].
\end{equation}
In other words, since $a_S$ and $b_S$ are real numbers, during the
passage through the boundary, the polar angle of the spin relative
to the axis along which the tilt occurs changes, while the azimuthal
angle perpendicular to the tilt direction remains unchanged.\par
The above calculations demonstrate that at the interface of the two
regions, the pseudo-spin direction remains in the $x-y$~plane, but
its orientation with respect to the $x$-axis in the $x-y$~plane
changes.\par
In the following, we examine quantum transport of the Dirac fermions
in normal and oblique incidence of the electron wavefunction on the
interface of two mediums as two separate cases.
\subsubsection{Normal incidence analysis}
Let us consider the case where an electron approaches to the
interface from the left side of the considered heterojunction. To
simplify our analysis, we will begin with the assumption of a
normally incident electron, which implies $k_y = 0$. We will also
consider that the Dirac cone in region $B$ is in a sub-critically
tilted phase $(0 < \zeta_B < 1)$. In this case, the wave functions
can be expressed as:
\begin{equation} \label{Eq21}
\psi(x) = \begin{cases} e^{ik_xx} \Big(\begin{array}{c} 1 \\  1
\end{array}\Big) + r e^{-ik_xx}
\Big(\begin{array}{c} 1 \\ -1  \end{array}\Big), & x < 0, \\t
\Big(\begin{array}{c} 1 \\  1  \end{array}\Big) e^{ik_xx}, & x > 0,
\end{cases}
\end{equation}
where the first row is assumed as the electron wave function in
region $A$ and the second row as the same in region $B$. $r$ and $t$
represent the reflection and transmission amplitudes, respectively.
By applying the boundary conditions given in Eq.~\eqref{Eq13} at the
interface assumed at $x=0$ and considering $\zeta_A = 0$, the
following coupled equations can be easily derived:
\begin{equation}\label{Eq22}
\begin{split}
1+r = (\sqrt{1+\zeta_B})t,\\
1-r = (\sqrt{1+\zeta_B})t,
\end{split}
\end{equation}
for which the solutions are
\begin{equation}\label{Eq23}
r = 0,\qquad\text{and}\qquad t = 1/\sqrt{1+\zeta_B}.
\end{equation}
We can also calculate the transmission probability $T$, which is
defined in terms of the incident and transmitted probability current
densities denoted as $J_i$ and $J_t$, respectively,
\begin{equation}\label{Eq24}
T = {J_t\over J_i} = {J_{Bx}\over J_{Ax}}.
\end{equation}
As discussed earlier, it is clear that the incident probability
current density $J_i$ is $J_{Ax} = \psi_A^\dagger(v \sigma_x) \psi_A
= v$ and the transmitted probability current density  $J_t$ is $
J_{Bx} = \psi_B^\dagger(v_x \sigma_x + v_t{\bf 1}) \psi_B = v $.
Therefore, if the electron arrives to the interface with normal
incidence, the probability of it passing through is 1, meaning $T =
1$. This is often referred to as the Klein tunneling
effect~\cite{Dombey01}, which this result shows it occurs even in
the presence of tilt in the energy band dispersion where $|t|\neq
1$.\par
It is interesting that if the traveling direction of the electrons
is reversed, so that they incident from the tilt medium to the
interface, there is still a possibility of complete tunneling.\par
It is worth noting that when $\zeta_B > 1$, it is impossible to
express the wave function for region $B$ as given in
Eq.~\eqref{Eq21} because a particle with a group velocity moving
from right to left cannot be found in this region. Additionally, if
a particle crosses the boundary, it cannot escape, resembling the
event horizon of a black hole. This scenario is interesting and
reminiscent of the behavior near a black hole's event horizon.
\begin{figure}[h!]
\begin{center}
\includegraphics[scale=0.54]{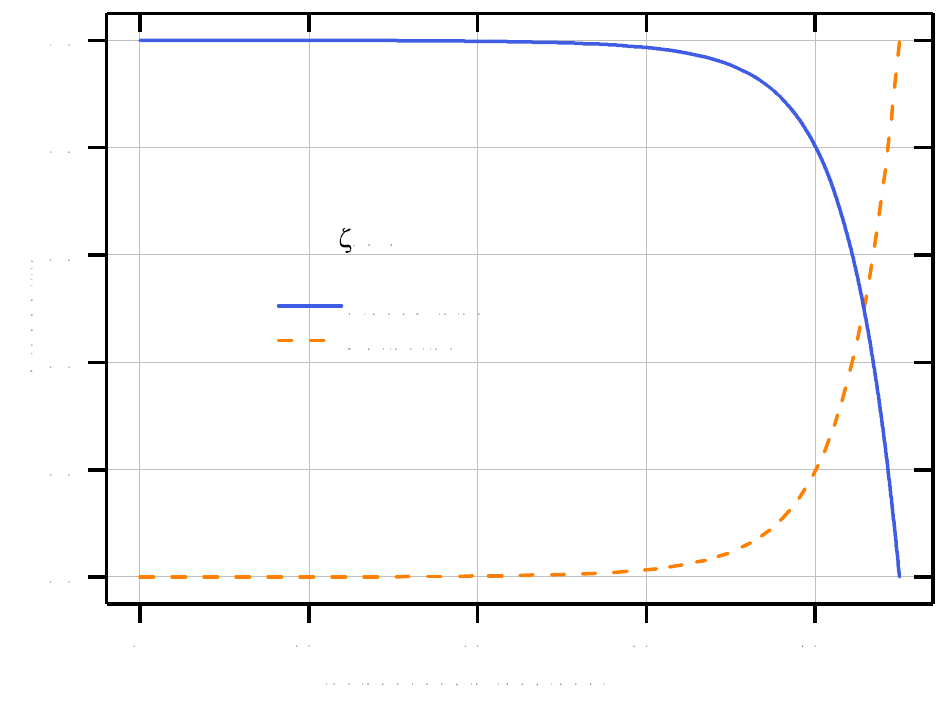}
\end{center}
\caption{ The probability of reflection and transmission for various
incident angles as the Dirac fermions enter the region with tilted
Dirac cones from the region without tilt (region $A$ to $B$).
\label{Fig03}}
\end{figure}
\subsubsection{Oblique incidence analysis}
Let us now explore the oblique incidence at the interface between
regions $A$ and $B$ as a second but more general case. We consider
an incoming electron with a given wavevector of ${\bf k}_i$, which
is reflected with wavevector ${\bf k}_r$ in region $A$, and is
transmitted into region $B$ with a wavevector of ${\bf k}_t$. This
scenario is more complex than the normal incidence case, as the wave
vectors and the angles of incidence, reflection, and transmission
are involved.\par
Referring to the angles illustrated in Fig.~\ref{Fig02}, we can
establish the wavevectors as ${\bf k}_\alpha = (k_{\alpha x},
k_{\alpha y}) = k_\alpha (\beta \cos \theta_\alpha, \sin
\theta_\alpha)$ for $\alpha = i,~r,~t$, in which indices $i$, $r$
and $t$ refer to the incident, reflected and transmitted waves,
respectively. $k_\alpha$ is the norm of ${\bf k}_\alpha$ and $\beta$
is $-1$ for the reflected and is $+1$ for the incident and
transmitted wave vectors. Given the conservation of energy $E$ and
momentum projection $k_y$, the component which is parallel to the
interface, we can immediately conclude $k_r = k_i$, $\theta_r =
\theta_i$, $k_{t y} = k_{i y}$,
\begin{equation}\label{Eq25}
\theta_t = \cot^{-1}{\Big({{-\zeta_B +
\sqrt{1-(1-\zeta_B^2)\sin^2{\theta _i} }\over
(1-\zeta_B^2)\sin\theta_i} \Big)}},
\end{equation}
and $k_{tx}$ reads as
\begin{equation} \label{Eq26}
k_{tx} =  {-\zeta_B E + \sqrt{\zeta_B^2 E^2 +(1- \zeta_B^2)
k_{ix}^2} \over 1-\zeta_B^2}.
\end{equation}
The associated wave functions to the incident, reflected and
transmitted electrons can be respectively expressed as:
\begin{equation}\label{Eq27}
\begin{split}
\psi_i({\bf r}) &= \Big(
\begin{array}{c} 1\\ e^{i \theta_i}
\end{array}\Big)  e^{i {\bf k}_i \cdot \bf{r}},\\
\psi_r({\bf r}) &= \Big(
\begin{array}{c}
1\\ -e^{-i \theta _r}
\end{array}\Big) e^{i {\bf k}_r \cdot \bf{r}},\\
\psi_t({\bf r}) &= \Big(
\begin{array}{c} 1\\ e^{i \theta_t}
\end{array}\Big)  e^{i {\bf k}_t \cdot \bf{r}}.
\end{split}
\end{equation}
These equations allows us to find the wave functions  in both
regions as
\begin{equation}\label{Eq28}\begin{split}
\psi_A ({\bf r}) & =    \psi_i ({\bf r})+ r \psi_r({\bf r}), \quad
\text{for} \quad x < 0,\\
\psi_B ({\bf r}) & = t \psi_t ({\bf r}),\qquad\qquad\ \ \text{for}
\quad x
> 0.
\end{split}
\end{equation}
Inserting these wavefunctions into Eq.~\eqref{Eq13}, the following
relations between the reflection and transmission amplitudes are
drivable:
\begin{equation}\label{Eq29}
\begin{split}
1 + r = a_B t + b_B t e^{i \theta _t},\\
e^{i \theta _i} - r e^{-i \theta _i} = b_B t + a_B e^{i \theta _t}.
\end{split}
\end{equation}
After some straightforward algebra, the solutions for $r$ and $t$
are obtained as:
\begin{equation}\label{Eq30}
\begin{split}
t & = {2\cos \theta_i \over 2 a_B \cos \theta_i + b_B(1 + e^{i (\theta_t - \theta_i)})},\\
r & = t(a _B+ b_B e^{i \theta_t}) - 1.
\end{split}
\end{equation}
For this case, the subsequent step involves calculating the
reflection and transmission coefficients, $R$ and $T$. This includes
calculating the ratios of $J_r/J_i$ and $J_t/J_i$. The perpendicular
probability flux in region $A$, $J_{Ax}$, can be expressed as:
\begin{equation}\label{Eq31}
J_{Ax} = v \psi_A^\dagger \sigma_x \psi_A = J_i + J_r,
\end{equation}
where $J_i$ and $J_r$ are contributions from the incident and
reflected wavefunctions, respectively. Subsequently, these
contributions are derivable as
\begin{equation}\label{Eq32}\begin{split}
J_i & = 2 v |1 + r|  \cos{(\theta_i - \delta_1)},\\ J_r & = 2 v |r(1
+ r)| \cos{(\theta_i + \delta_2)},
\end{split}
\end{equation}
where, $\delta_1$ and $\delta_2$ are the phases associated with
expressions $1 + r$ and $r^*(1 + r)$, respectively. Here the
asterisk refers to complex conjugate.\par
Similarly, the probability flux transmitted perpendicular, $J_t$,
reads
\begin{equation}\label{Eq33}
J_t = v \psi_B^\dagger (\sigma_x + \zeta_B {\bf 1}) \psi_B = 2|t|v
(\zeta_B+\cos\theta_t).
\end{equation}
Consequently, $R$ and $T$ are determined as
\begin{equation}\label{34}\begin{split}
R & = {J_r\over J_i} = {|r|\cos{(\theta_i+\delta_2)} \over
\cos{(\theta_i-\delta_1)}},\\ T & = {J_t \over J_i} =
{|t|^{2}[\cos{\theta_t+\zeta_B}]\over
2|1+r|\cos{(\theta_i-\delta_1)}}.
\end{split}
\end{equation}
In Fig.~\ref{Fig03}, the behavior of both the reflection and
transmission probabilities, $R$ and $T$, as functions of incident
angle $\theta_i$ is visually represented. The phenomenon of Klein
tunneling remains prominent for incident angles less than
$50^\circ$, with $T$ reaching nearly~$1$. However, as $\theta_i$
exceeds $50^\circ$, the probability of reflection becomes notably
significant. It is worth noting that in this figure, the tilt
parameter is assumed to be $\zeta_B = 0.4$.  The change in these
probabilities with the incident angle $\theta_i$ and the tilt
parameter $\zeta_B$ provides valuable insights into the quantum
behavior of the system.\par
\begin{figure*}[t!]
\begin{center}
\includegraphics[scale=0.5]{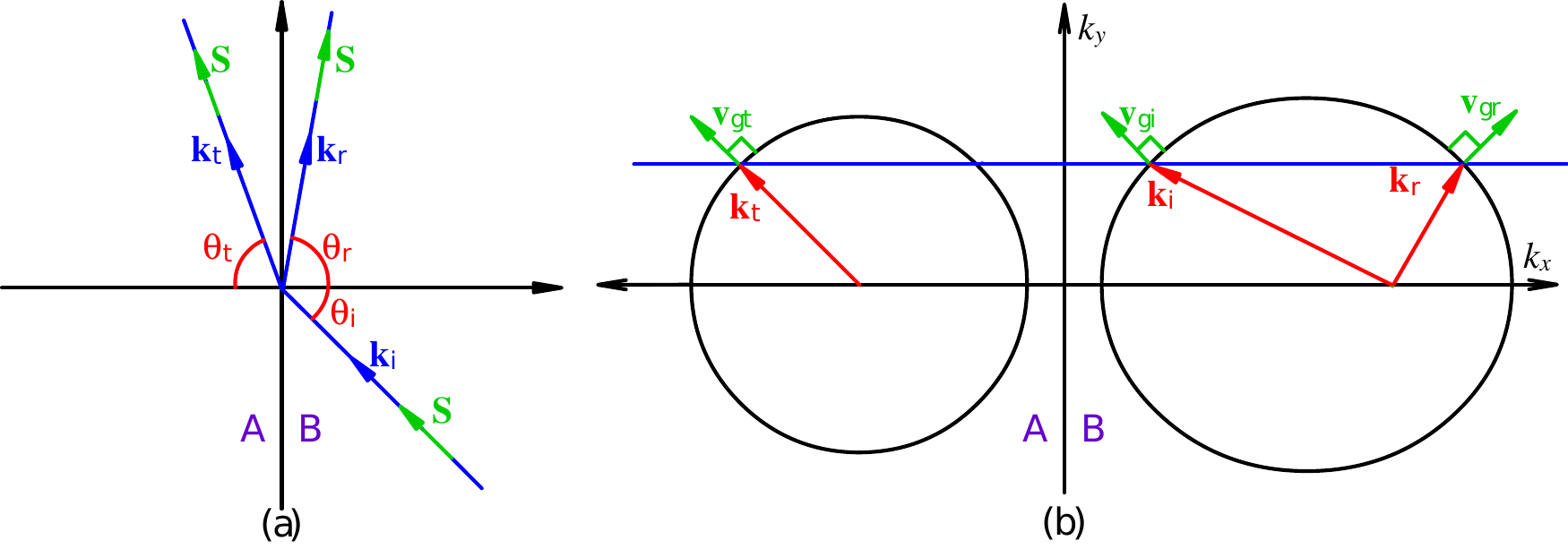}
\end{center}
\caption{Same as figure~\ref{Fig02}, but for incidence of Dirac
fermions from the region with tilted cones to the region without
tilting \label{Fig04}}
\end{figure*}
Before closing this subsection, it is interesting to ask what will
happen if the incident direction is reversed from right to left on
the hetrojunction. To explore the scenario of transmission from the
tilted side to the untilted one (reversing the incident direction),
we need to consider the corresponding angles and wavevectors, which
are depicted in Fig.~\ref{Fig04}. By applying the conservation laws
discussed earlier, we can numerically determine  $\theta_t$ using
the following relation:
\begin{equation}\label{Eq35}
\theta_i = \cot^{-1}\Big({-\zeta_B -
\sqrt{1-(1-\zeta_B^2)\sin^2{\theta_t}} \over
(1-\zeta_B^2)\sin\theta_t}\Big),
\end{equation}
and then use the result to determine $\theta_r$ as
\begin{equation}\label{Eq36}
\theta_r  = \cot^{-1}\Big({-\zeta_B +
\sqrt{1-(1-\zeta_B^2)\sin^2{\theta_t}}\over(1-\zeta_B^2)\sin\theta_t}\Big).
\end{equation}
This procedure comes from the fact that obtaining the closed
analytical form for $\theta_t$ from Eq.~\eqref{Eq35}, is complicated
and difficult.\par
\begin{figure}[t!]
\begin{center}
\includegraphics[scale=0.5]{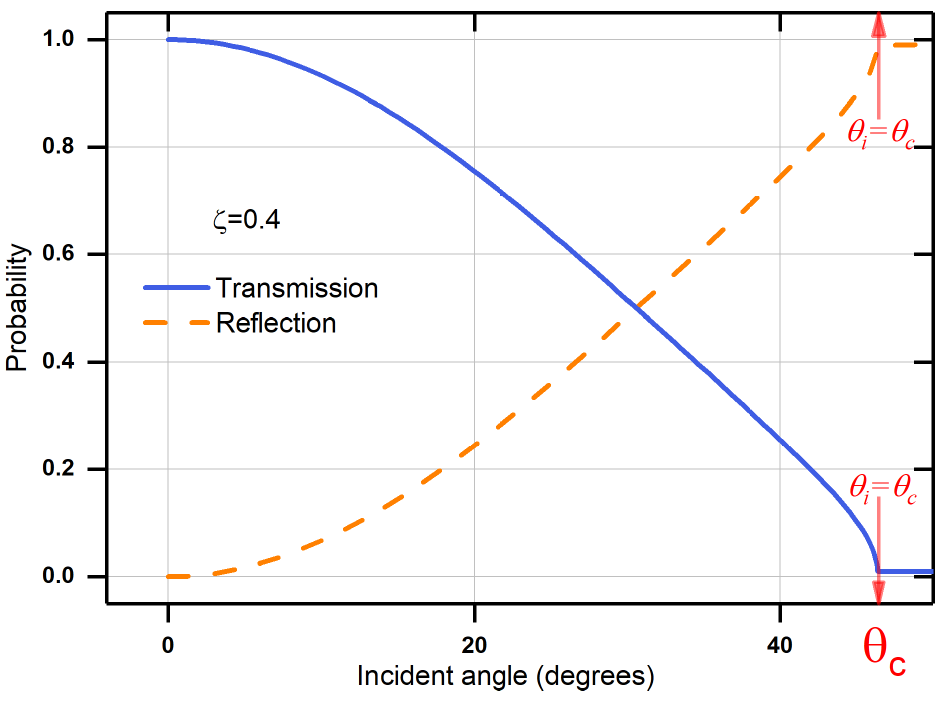}
\end{center}
\caption{Same as figure~\ref{Fig03}, but for incidence of Dirac
fermions from the region with tilted cones to the region without
tilting. In this case there exists a critical angle, $\theta_c\simeq
46.4^\circ$, for which the fermion waves are totally reflected into
the tilted material.\label{Fig05}}
\end{figure}

The previous analysis can be extended to determine the reflection
and transmission probabilities for this case too. For $\zeta_B =
0.4$, Fig.~\ref{Fig05} displays the graphical representations of
computed $R$ and $T$ as functions of the incident angle, $\theta_i$.
The overall behavior is both qualitatively and quantitatively
different from the previous case. The significant point that we get
from the comparison of Figs.~\ref{Fig03} and~\ref{Fig05} is that the
behavior of Dirac fermions at the boundary of two materials with
different tiling depends on the incident direction. In the incidence
from the untilted region to the tilted region, for an angular range
of 0 to 50 degrees, the probability of the quantum Klein tunneling
is perfect, but it is not the case in inverse. Also, another
significant difference that can be seen in Fig.~\ref{Fig05} is the
existence of a  critical incident angel for the case in which Dirac
fermions entering from the tilted region to the region with upright
cones. In Fig.~\ref{Fig05}, this angle is shown by~$\theta_c$. In
optics terminology, this angle refer to the specific angle of
incidence that results in an angle of refraction of $90^\circ$,
beyond which total internal reflection occurs. In fact, for
incidence angles greater than ~$\theta_c$, entry into region $A$
becomes prohibited. This angle is obtained by setting $\theta_t$
equal to $90^\circ$ in Eq.~\eqref{Eq35}, for which the result is
$\theta_c = \cot^{-1}\big( -2 \zeta_B/(1-\zeta_B^2)\big)$. The
critical angle $\theta_c$, which introduces an additional constraint
on the system is a key feature of the considered configuration, and
the figure effectively conveys this information. For the case
corresponding to Figs.~\ref{Fig02} and~\ref{Fig03}, this critical
angle is absolutely absent.\par
\begin{figure*}[t!]
\begin{center}
\includegraphics[scale=0.8]{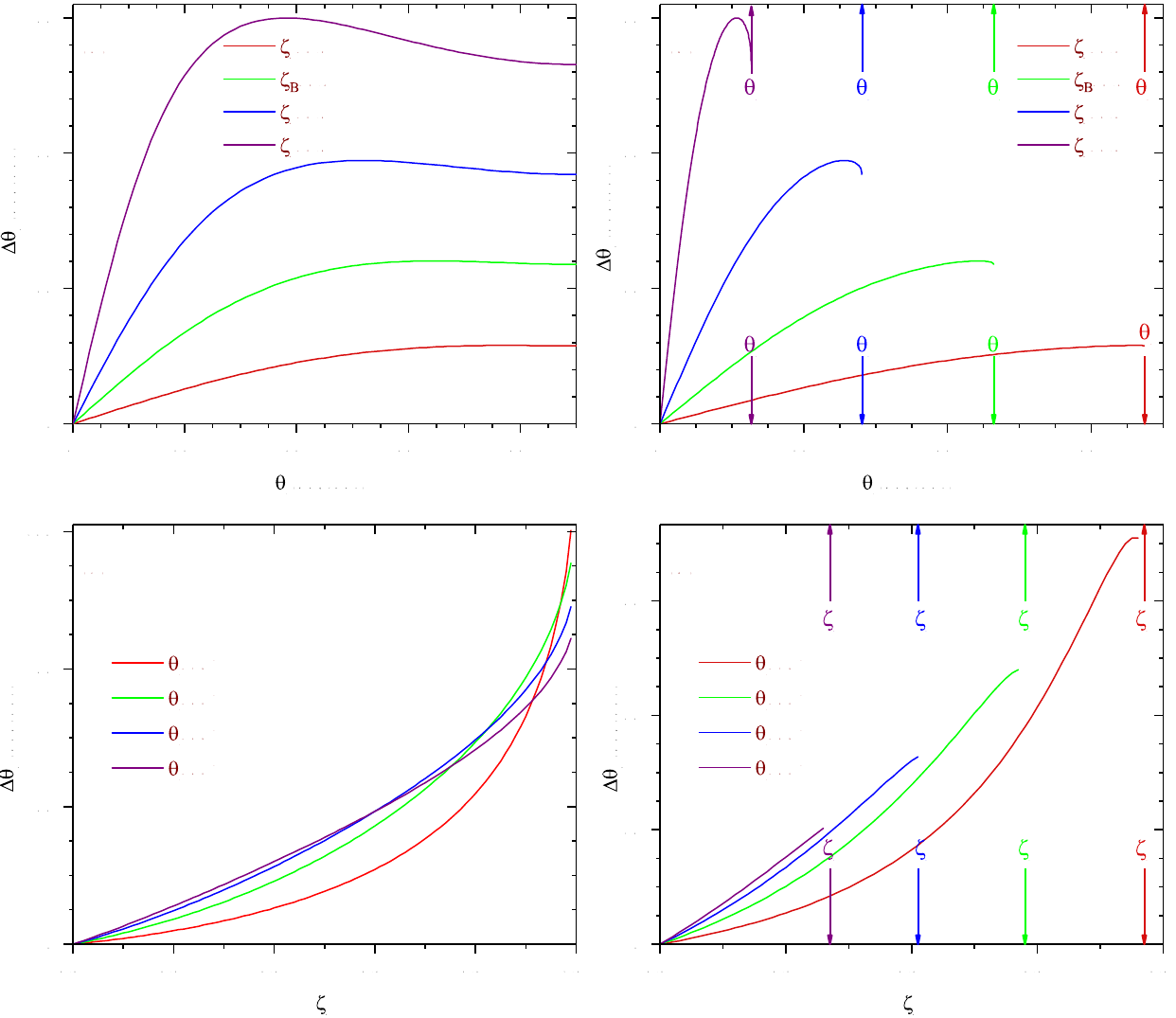}
\end{center}
\caption{The change in angular orientation of the pseudospin of
massless Dirac electrons, denoted as $\Delta\theta_S$, as they cross
the interface between two Dirac materials having different tilt
parameters: The left column is for the setup presented in
Fig.~\ref{Fig02} and the other for the same in Fig.~\ref{Fig04}. The
first row displays $\Delta\theta_S$ as a function of~$\theta_i$ for
some fixed values of $\zeta_B$ and the second row demonstrates it as
a function of~$\zeta_B$ but for several fixed values of~$\theta_i$.
In the right column the critical incident angle and tilt parameter
are denoted by $\theta_c$ and $\zeta_c$, respectively.
\label{Fig06}}
\end{figure*}

Another important item to address in this subsection is the angular
amount of the rotation of the pseudospin of the massless Dirac
fermions in crossing the boundary of the Dirac materials with
different tilts. In Fig.~\ref{Fig06}, a comprehensive analysis of
this significant issue has been conducted. In this figure the left
column is for crossing from~$A$ to~$B$ and the other is for the
reversed crossing. Also the first row is produced to show the amount
of the spin rotation, $\Delta \theta_S = \theta_S -\theta$, as a
function of the incident angle, $\theta_i$, for some fixed values of
$\zeta_B$.  The second row demonstrates $\Delta\theta_S$ as a
function of tilt parameter $\zeta_B$, for some fixed values of
$\theta_i$. All the graphs are plotted by employing
Eq.~\eqref{Eq20}. As is seen from the figure, in both cases,
$\Delta\theta_S$ is almost an increasing function of $\theta_i$
and/or $\zeta_B$. In panel (a), The slope of the changes is high at
the beginning, but at larger angles, this slope is adjusted and even
turns slightly downward. For entering from $B$ to $A$, as it has
been explained in the discussion provided on Fig.~\ref{Fig05}, for
each fixed value of $\zeta_B$, there exists a critical incident
angle at which the electron wave is totally reflected into the
tilted medium. These critical angles are marked with arrows in
panel~(b). In panels~(c)~and~(d), the slope of changes is absolutely
positive and increasing.  Panel~(d) shows that for any given
incident angle, there exists a certain value of the tilt parameter,
beyond which the electronic wave is entirely reflected into the
medium~$B$. For different values of $\theta_i$, these critical
values are labeled with $\zeta_c$ and are shown with arrows in this
panel. In the first row, $\theta_i$ has changed from zero to $\pi/2$
or to $\theta_c$. If the interval of $\theta_i$ changes is taken to
be symmetrical, for example $[-\pi/2,+\pi/2]$ or
$[-\theta_c,\theta_c]$, the graphs will be symmetrical with respect
to the vertical axis, that is, $\Delta \theta_s$ is an even function
of $\theta_i$. Of course, this feature is confirmed by current
calculations and it can be found out from Figs.~\ref{Fig02}
and~\ref{Fig04}. It is worthy to mention that, in all the considered
cases, there is no observation of a change in the sign of the
pseudospin of the massless electrons.
\subsection{Double interface heterojunction\label{Sec02SubC}}
\begin{figure}[t!]
\begin{center}
\includegraphics[scale=0.35]{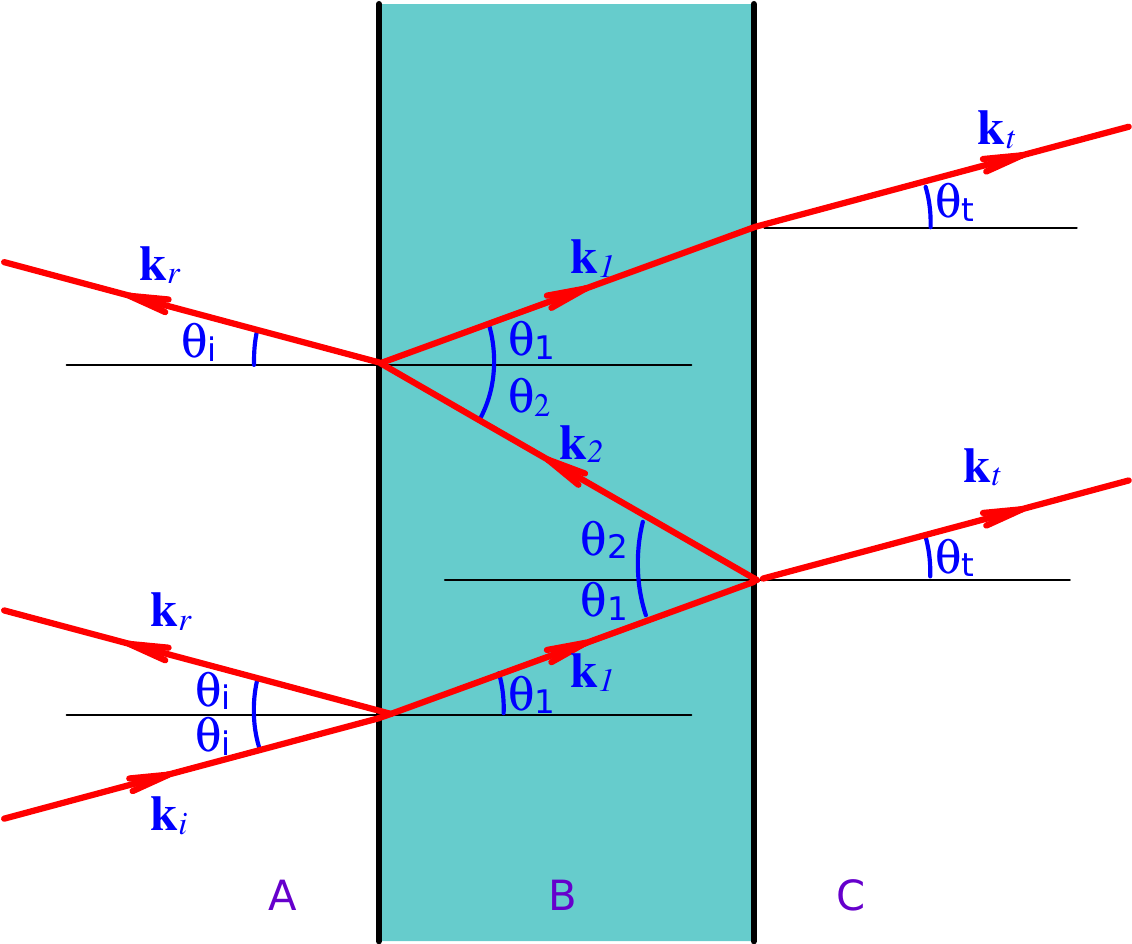}
\end{center}
\caption{Incident Dirac fermion wave with a wavevector of ${\bf
k}_i$ strikes the slab of an tilted fermion Dirac material (region
B) placed between two untilted Dirac mediums (regions A and C),
giving rise to the reflected and transmitted Dirac fermion waves
with wave vectors of ${\bf k}_r$ and ${\bf k}_t$, respectively.
\label{Fig07}}
\end{figure}

\begin{figure}[h!]
\begin{center}
\includegraphics[scale=0.55]{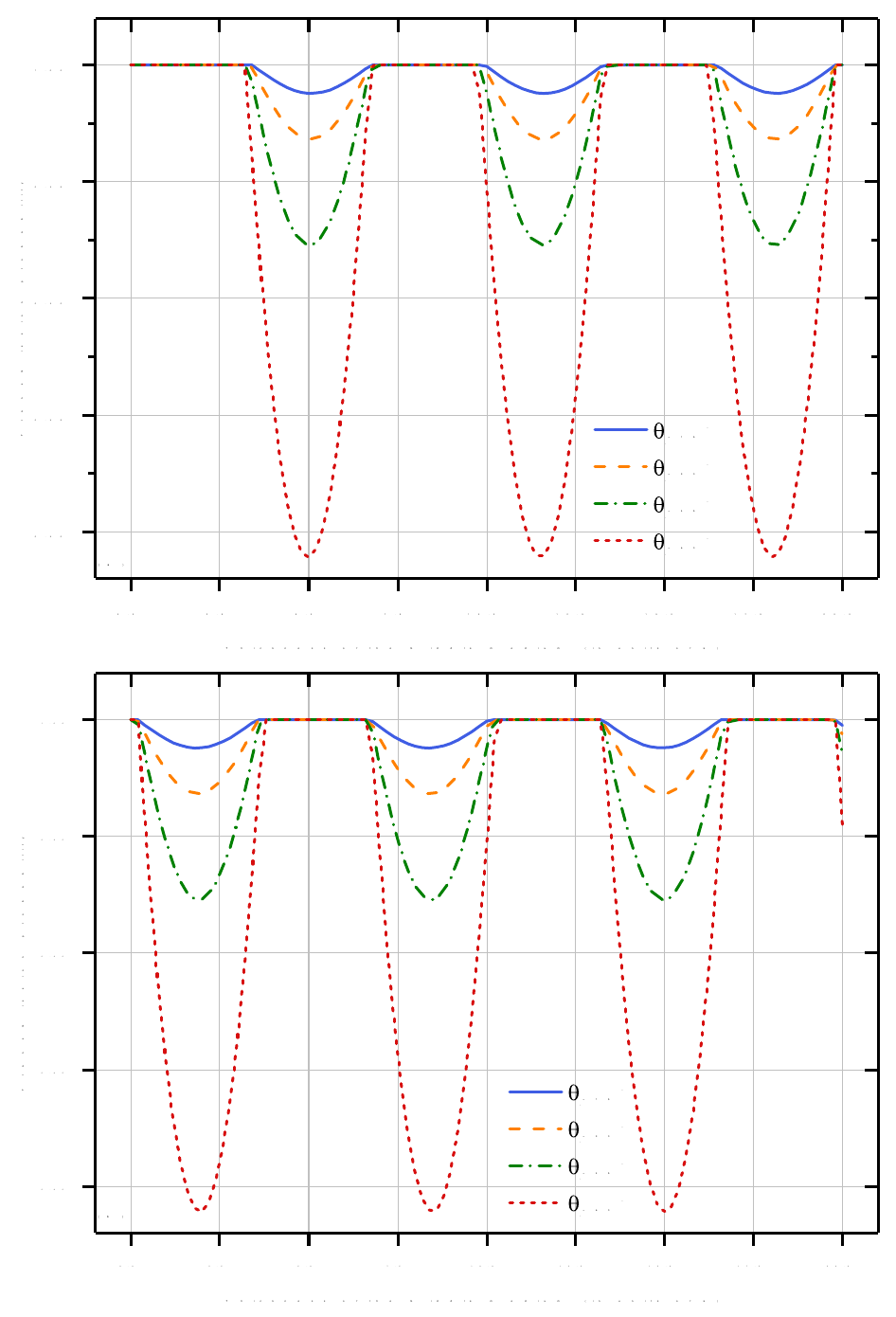}
\end{center}
\caption{Quantum tunneling probability of Dirac fermions within a
slab with thickness $\ell$ for configuration illustrated in
Fig.~\ref{Fig07} in two different situations: (a) Forward
traveling~(left to right) and (b) Reverse traveling (right to left).
The transmission probability for various incident angles is plotted
against the slab's thickness, $\ell$. In both cases
$\zeta_B=0.4$.\label{Fig08}}
\end{figure}

Now, let us explore a heterojunction with double interfaces, as
depicted in Fig.~\ref{Fig07}. This setup comprises three regions,
denoted as $A$, $B$, and $C$. Regions $A$ and $C$ contain untilted
Dirac cones, while the middle region, region $B$, features a tilted
Dirac cone, represented by $\zeta_B$. The middle slab's thickness
containing the subcritical Dirac cones is denoted as $\ell$. For the
specified three-medium structure, depicted in Fig.~\ref{Fig07}, the
corresponding wave functions for the regions can be written as
\begin{equation}\label{Eq37}
\psi(x) = \begin{cases} \Big( \begin{array}{c} 1\\ e^{i \theta_i}
\end{array} \Big) e^{i {\bf k}_i \cdot \bf{r}} + r \Big(
\begin{array}{c}1\\ e^{i \theta_r}\end{array} \Big) e^{i {\bf k}_r \cdot \bf{r}
}, & \text{for } x < 0, \\ D_1 \Big( \begin{array}{c} 1\\
e^{i\theta_1} \end{array} \Big) e^{i {\bf k}_1 \cdot \bf{r}}+ D_2
\Big(
\begin{array}{c} 1\\ e^{i \theta_2}
\end{array} \Big) e^{i {\bf k}_2 \cdot \bf{r}
}, & \text{for } 0 < x < \ell, \\ t \Big( \begin{array}{c} 1\\ e^{i
\theta_t} \end{array} \Big) e^{i {\bf k}_t \cdot \bf{r} }, &
\text{for } x > \ell, \end{cases}
\end{equation}
where $D_1$ and $D_2$ are two constant coefficients, ${\bf k}_1$ and
${\bf k}_2$ are the wave vectors of the  traveling and reflected
waves in the middle region, and angles $\theta_1$ and $\theta_2$
indicate the orientation of these wave vectors with respect to the
direction of tilting. The indicated angles along with the components
of the wave vectors such as $k_{1x}$, $k_{2x}$, and $k_{tx}$ can be
obtained using the energy and momentum conservation laws as in the
previous section. These results obtained from the mentioned
procedure are
\begin{equation}\label{Eq38}
\theta_1 = \cot^{-1}{\Big({-\zeta_B - \sqrt{1-(1-\zeta_B^2)
\sin^2{\theta_t}} \over (1-\zeta_B^2)\sin\theta_t}\Big)},
\end{equation}
\begin{equation}\label{Eq39}
\theta_2 = \cot^{-1}{\Big({-\zeta_B + \sqrt{1-(1-\zeta_B^2)
\sin^2{\theta_t}} \over (1-\zeta_B^2)\sin\theta_t} \Big)},
\end{equation}
\begin{equation}\label{Eq40}
k_{Ax} = {-\zeta_B + \sqrt{1-(1-\zeta_B^2) \sin^2{\theta_t}} \over
1-\zeta_B^2},\qquad\qquad
\end{equation}
\begin{equation}\label{Eq41}
k_{Bx} = {-\zeta_B - \sqrt{1-(1-\zeta_B^2) \sin^2{\theta_t}}\over
1-\zeta_B^2},\qquad\qquad
\end{equation}
and
\begin{equation}\label{Eq42}
k_{tx} = {E\over v_x}\cos \theta_t.
\end{equation}
To obtain the coefficients $D_1$, $D_2$ and the reflection and
transmission amplitudes $r$, and $t$, we need to employ the boundary
condition given in Eq.~\eqref{Eq13} on both interfaces assumed at $x
= 0$ and $x = \ell$. The corresponding equations read
\begin{equation}\label{Eq43}
\psi_A(0) = M_B \psi_B(0),
\end{equation}
\begin{equation}\label{Eq44}
M_B \psi_B(\ell) = \psi_C (\ell).
\end{equation}
After performing the necessary calculations, the reflection and
transmission probabilities, $R$ and $T$, can be derived. Although
the closed but complex forms of these expressions are not provided
here, they can be utilized to examine the behavior of Dirac fermions
in confronting with the interfaces. As an example, the changes of
$T$ in terms of the middle slab's thickness, for two cases of
incidence from left to right and incidence from right to left are
visualized in Fig.~\ref{Fig08}. In this figure, panels~$(a)$ and
$(b)$ illustrate the behavior of the transmission probability for
the specified cases, respectively. As is seen, the plots exhibit two
different but nearly opposite behaviors. In other words, in regions
where the probability of transmission from left to right is nonzero,
the probability of transmission from right to left is perfect and
vice versa. This antisymmetry is so clear that the graphs can be
considered nearly complementary to each other. Another important
point is that the transmission probability is a periodic function of
the thickness of the middle region with the subcritical tilted Dirac
cones, and the period of oscillation depends on the tilt parameter
of this medium, $\zeta_B$. Also, for certain thicknesses of the
middle slab, the probability of quantum tunneling of the Dirac
fermions inside the slab is perfect. For both considered cases,
these thicknesses are in some continuous and wide ranges. Another
point is that the thicknesses in which the Klein tunneling occurs
are independent of the incidence angles, so that for all the angles
shown in the figure in both cases, Klein tunneling occurs in the
same range of thickness. With the features listed above, it seems
that tuning the length of the middle slab to observe this phenomenon
experimentally probably is not so difficult.\par
As expected and  the above studies confirm, when Dirac fermions
cross the interface between two materials with different tilts,
their behavior can indeed be different depending on the direction of
traversal. The resulting phenomena can have implications for novel
electronic devices and fundamental research in condensed matter
physics.
\section{Summary and Conclusion\label{Sec03}}
In this study, we have thoroughly investigated the profound
implications of tilted Dirac cones on the quantum transport
properties of two-dimensional (2D) Dirac materials. Our research has
focused on materials with tilted Dirac cones, where the anisotropic
and tilted nature of the cones introduces additional complexity and
richness to their electronic properties. The investigation began by
considering a heterojunction of 2D Dirac materials, where electrons
undergo quantum tunneling between regions with upright and tilted
Dirac cones. We have provided a comprehensive theoretical
investigation into the impact of tilted Dirac cones on electron
transmission and pseudospin dynamics in 2D materials. Our study has
revealed several key findings. We have derived boundary conditions
governing reflection and transmission between two regions
characterized by different tilts, and quantified the spin rotation
at the interface. Furthermore, we have investigated the
probabilities of electron reflection and transmission during the
transition from a non-tilted region to a tilted one. Our results
have demonstrated that the probability of massless electron
transmission through a tilted region exhibits a periodic dependence
on the width of an intermediate region, with this periodicity being
independent of the incident angle. Importantly, we have observed
that, for sufficiently large region widths, the transmission
probability approaches unity. Additionally, we have highlighted the
asymmetry of transmission probabilities between left-to-right and
right-to-left directions.\par
Electrons transmitted through an interface across which the tilting
parameter abruptly changes undergo a rotation of its pseudospin. In
the case of spin-orbit coupled Dirac cones, such as those in the
surface of topological insulators, this would imply that the
interface is capable of rotating the incoming spins. Although in the
present study we assumed that the tilt parameter changes abruptly
(and so does the pseudo-spin rotation), in a generic setting where
the tilt parameter changes smoothly it is again expected to have a
smooth rotation of spin orientation. Since the tilt parameter is a
proxy for spacetime
metric~\cite{Volovik01,Volovik02,Volovik03,Mola01,Mola02} this can
be considered as an example of a \emph{gravitomagnetic} effect,
where spatial variation of certain metric entries may have an effect
that effectively looks like a ''magnetic field"~\cite{Ryder01}.\par
The implications of our research extend to the ongoing efforts to
manipulate and precisely tune the tilt in Dirac materials.
Furthermore, our study has shed light on the broader implications of
tilted Dirac cones, particularly in less symmetric Dirac materials,
where the Coulomb interaction can give rise to even more exotic
phenomena. Moreover, we have delved into the theoretical exploration
of the excitonic transition in 2D tilted cones to understand the
electron-hole pairing instability as a function of tilt, providing
valuable insights into the chiral excitonic instability of such
systems. In conclusion, our study has significantly advanced the
understanding of quantum transport phenomena in 2D materials with
tilted Dirac cones. The insights gained from this research have the
potential to inform the development of novel quantum devices and
pave the way for further theoretical and experimental investigations
into the unique electronic properties of materials with tilted Dirac
cones.
\section*{Acknowledgment}
The first author, RAM, would like to acknowledge the office of
graduate studies at the University of Isfahan for their support and
research facilities. Also, the forth author, MA, acknowledges the
support received through the Abdus Salam (ICTP) short visit program.

\end{document}